# How do activity-end trip characteristics affect the choice for shared micromobility?

A latent class choice modelling approach for train station egress trips in the Netherlands

Nejc Geržinič[1]*, Mark van Hagen[2], Hussein Al-Tamimi[3], Dorine Duives[4], Niels van Oort[4]

[1] Postdoctoral researcher, Department of Transport & Planning, TU Delft, Delft, Netherlands
[2] Principal consultant, Dutch Railways (NS), Utrecht, Netherlands
[3] Propositions & Partnerships, Dutch Railways (NS), Utrecht, Netherlands
[4] Associate professor, Department of Transport & Planning, TU Delft, Delft, Netherlands
* corresponding author

## Abstract

Access/egress travel to train stations continues to pose a significant barrier to increasing the number of train travellers. Shared micromobility (SMM), including bicycles, e-bikes, steps and mopeds, is often cited as a prominent solution, especially for the activity-end of the trip chain. Using a stated preference survey, we analyse activity-end mode-choice preferences for SMM, walking and public transport (PT) among the Dutch population. By means of a latent class choice model, we uncover three user groups with respect to activity-end mode choice behaviour. The largest (58%) are *Multimodal sharing enthusiast*, who choose based on the trade-offs between various travel characteristics, while not having strong modal preferences. They are the most open, ready and able to use SMM. *Sharing hesitant cyclists* (16%) have a strong preference for cycling and while they are open to using SMM , they may not feel themselves ready, stating that use of SMM can be difficult and dangerous. *Sharing-averse PT users* (27%) are most likely to use PT and avoid SMM as they find it too difficult and dangerous to use. The high preference to walk for short egress distances reaffirms the need for transit-oriented development policies. For longer egress distances, PT should be the primary focus at stations in high-density areas with high demand, where high frequencies and dense networks are justified, while stations in lower demand areas are better served by SMM. Providing multiple SMM options would result mainly in competition for the same travellers.

## Keywords

Access/egress travel, Walking, Cycling, Shared micromobility, Stated choice experiment, Latent class choice model

## 1  Introduction

Achieving a sustainable mobility sector remains a critical global objective in the effort to combat climate change and reduce greenhouse gas emissions. In Europe, the transport sector stands out as the only sector where emissions have increased since 1990 (European Environment Agency, n.d.), further highlighting the need to address its environmental impact. The private car in particular remains a key issue to address, representing the largest share in the modal split (Prieto-Curiel & Ospina, 2024) and the substantial externalities (GHG emissions, noise, traffic safety, space consumption,…) associated with its use (Brand et al., 2021; Gössling et al., 2019; McLaren et al., 2015). As a large share of car trips are short, often less than 5km (de Graaf, 2015; Mackett, 1999; US Department of Energy, 2022), shifting them to active mobility (walking, cycling and other forms of micromobility) is the most sustainable solution (Brand et al., 2021; Gössling et al., 2019; McLaren et al., 2015).

For travel distances beyond ~5km, often considered as the upper threshold for cycling (Jonkeren & Huang, 2024; Kager et al., 2016; Keijer & Rietveld, 2000; Martens, 2004; Rietveld, 2000), (rail-based) public transport proves to be the most efficient and sustainable alternative for the private car (Brand et al., 2021; McLaren et al., 2015). Although achieving a high average speed between stations, the door-to-door travel speed of public transport is highly impacted by access/egress travel (Krygsman et al., 2004). This refers to travel to a public transport stop before the trip, and from the stop to the final destination. While faster modes can be used here as well (i.e., the car), the majority of access and egress trips are performed by active modes (walking and cycling) (Azimi et al., 2021; Keijer & Rietveld, 2000; Ton et al., 2020). Because of this, access and egress legs together can account for up to 50% of the door-to-door travel time (Krygsman et al., 2004), despite constituting only a small fraction of the distance. It also means that improving these trip legs can provide substantial benefit to the overall travel experience.

In countries with high bicycle usage, using it for access/egress trips is particularly popular. As highlighted by Kager et al. (2016), the bicycle-train combination has multiple benefits, for the traveller and society. The bicycle effectively complements the higher travel speed of trains by vastly improving accessibility of train stations, increasing the range and thus the number of people that can reach a train station in a reasonable time. In the Netherlands, for example, 19% of the population live within walking distance (1km) of a train station, while 69% live within cycling distance (5km) (Kager et al., 2016). A recent study by Jonkeren & Huang (2024) shows there is still potential for the bicycle-train combination to attract more car users. They conclude that up to 3.4% of trips or 7.8% of the travelled distance could be shifted from car to bicycle-train. While this may only represent a small share of car trips, it would result in a doubling of train travellers. Assuming a maximum cycling distance of 8km instead of 5km, this shift could increase to 4.9% or 11% of trips and travelled distance, respectively (Jonkeren & Huang, 2024). While such a cycling distance may be beyond the range of what most regular travellers are willing to cycle, electrified modes (e-bikes, e-scooters,…) may be able to further increase station catchment areas.

A substantial barrier to achieving this shift to rail travel lies on the activity-side of the trip (Shelat et al., 2018), often also referred to as the egress side in literature. Most travellers do not have a private bicycle available on both sides of the trip. Even in a country like the Netherlands, with bicycles outnumbering people, only ~10% of regular train commuters have a 'second' bicycle on their activity side (Schakenbos & Ton, 2023). This means the majority of travellers are limited to walking, taking public transport, or using a car (shared car, taxi or being picked up by someone). One possible solution for the activity end could be shared (micro)mobility (SMM). Although present for years, SMM truly took off in the last decade, with the advent of the internet and smartphones. A successful example of this is the OV-fiets (PT-bicycle) bicycle-sharing system in the Netherlands, a two-way station-based SMM system that started in 2008 and has since expanded to over 300 stations, with 5.9 million annual trips (NS Annual Report 2023, 2024).

As SMM became commonplace around the world, a growing body of research has emerged on the topic, examining its impact on travel behaviour, the environment and other modes, with many focusing especially on the impact on public transport. Abduljabbar et al. (2021) carried out an extensive overview of studies on the topic. In general, micromobility is found to contribute to sustainability goals in transportation, such as reducing congestion and emissions. They also highlight improved accessibility of public transport as a major benefit, while also pointing out that this particular topic still requires additional research. This research gap is also supported by Zhu et al. (2022) and Esztergár-Kiss & Lopez Lizarraga (2021). Studies on the interaction between micromobility and public transport show mixed outcomes, with some suggesting SMM is replacing (local) public transport trips (Badia & Jenelius, 2023; de Wit, 2023; Esztergár-Kiss & Lopez Lizarraga, 2021; Montes et al., 2023; Nikiforiadis et al., 2021; Reck & Axhausen, 2021; van Marsbergen et al., 2022; Wang et al., 2022), while others conclude that the impacts are limited (de Bortoli, 2021; Mehzabin Tuli et al., 2021).

With respect to integrating SMM with public transport, a review by Oeschger et al. (2020) finds that the focus tends to be centred around rail-based or higher speed public transport, with the analysis being based both on revealed (smartcard) and stated (survey) data. The main gaps they outline are a modal shift potential and limited knowledge of new electric SMM services. In terms of improving accessibility to public transport, Liu & Miller (2022) find that micromobility does improve accessibility, whereas Nawaro (2021), Ziedan, Darling, et al. (2021) and Ziedan, Shah, et al. (2021) all conclude that the impact is negligible. It should be noted that all four studies used RP data from systems that are not integrated with public transport but operated independently. Studies making use of SP data (Montes et al., 2023; Oeschger et al., 2023; Stam et al., 2021; Torabi K et al., 2022; van Kuijk et al., 2022; Yan et al., 2023) tended to focus on the access and/or egress leg, comparing the different micromobility modes for the first/last mile and different types of public transport as the main mode. The consensus among all is that newer SMM (mopeds and scooters) tends to be less preferred, with walking generally having the highest overall preference. They also agree that previous experience with shared modes is one of the strongest predictors of a more positive perception of SMM and a higher probability of their use.

Among the listed studies, van Kuijk et al. (2022) and Stam et al. (2021) were among the ones to include the largest numbers of SMM alternatives. However, the focus of van Kuijk et al. (2022) is primarily in urban areas, studying the relation between SMM and local PT like buses and trams, whereas Stam et al. (2021) carried out a simplified stated choice experiment without attributes to describe the individual alternatives. Studies analysing the relation between SMM and longer distance rail-based public transport (Oeschger et al., 2023; Torabi K et al., 2022; Yan et al., 2023) on the other hand included fewer SMM alternatives. None of the studies also included SMM system design characteristics like parking and rental type, the understanding of which is also mentioned as a gap in literature by van Waes et al. (2018).

Having identified the aforementioned research gaps, the contributions of this research are as follows. We (1) expand on the knowledge of the role of a wide variety of SMM modes for rail-based PT egress, (2) assess how trip circumstances, such as train trip length and trip purpose, affect egress mode choice and (3) analyse how travellers value SMM system design characteristics. This will allow us to (4) test travellers' sensitivity to individual attributes, and (5) evaluate the impacts of introducing new SMM services.

The rest of this paper is structured as follows: the survey design and modelling approach are outlined in Section 2. The modelling results are presented in Section 3 with an in-depth interpretation and a discussion of policy implications following in Section 4. Lastly, the conclusion, limitations and recommendations for future research steps are highlighted in Section 5.

## 2 Methodology

In this section, we present the approach undertaken in this research. We start by describing the survey design process used to capture behavioural information of individuals in Section 2.1. Next, the model formulation and estimation approach is outlined in Section 2.2. Finally, the data collection process, including data processing and sample representativeness are discussed in Section 2.3.

### 2.1 Survey design

To analyse the potential of SMM as an egress mode to train travel, we employ a stated preference (SP) data collection approach. While certain SMM modes, namely the 'regular' shared bicycle, are widely available around the Netherlands at train stations, this is not the case for other SMM modes, meaning that the availability of revealed preference data is likely limited. Additionally, to have full control over the attributes and to test the impact of individual characteristics, we opt for an SP approach as opposed to a revealed preference (RP) data collection approach.

The SMM modes we wish to test are the bicycle, e-bike, e-scooter and e-moped. In the Netherlands, the scooter (standing) is referred to as a step, while the moped (sitting) is called a scooter. To avoid confusion, we forego the use of the word scooter and use "step" when referring to the standing scooter and "moped" when referring to the sitting scooter. In addition to SMM modes, we include walking and local public transport like bus, tram or metro (BTM), as these are the most common egress modes in the Netherlands (Kennisinstituut voor Mobiliteitsbeleid, 2023).

The alternatives are described by a set of attributes. Time- and cost-related parameters tend to be the most influential for travellers and are key for analysing the trade-off behaviour (Montes et al., 2023; Oeschger et al., 2023; van Kuijk et al., 2022). We specify three common different travel time components (Wardman, 2004): (1) in-vehicle time refers to the travel time during the egress trip in or on the vehicle of choice, (2) walking time includes the walking distance from the train platform to the egress vehicle or in the case of walking as an egress mode, it is considered as the time needed to walk all the way to the destination and (3) waiting time, which only applies to the public transport mode, reflecting the time between arriving to the vehicle/platform and it's departure. Additionally, we specify two SMM design-related attributes, as outlined by van Waes et al. (2018), namely (1) how the vehicles are parked and (2) what kind of rental scheme is implemented.

As we wish to test the potential of SMM for different egress distances, we specify three different distance classes, namely 1km, 4km, and 7km. The attribute levels of in-vehicle time and travel cost depend on the egress distance, while other attributes do not. For the in-vehicle time, we use a pivot design based on average travel speeds to determine the values. We specify three levels, with the middle level being for the exact distance, while the lower and upper level are 25% below or above that distance respectively. The values are then rounded to minutes for the sake of clarity and simplicity. Cost is calculated based on current pricing (Check, 2024; NS, 2024; U-OV, 2025) of the modes for the three distance classes, from which the three levels are extracted. The lowest cost level is always "free" as we wish to also test the impact of the including the egress mode in the train ticket.

For the distance-independent attributes, we set attribute values that enable us to test a wide range of possible future scenarios and service designs. For walking and waiting time, we use levels that correspond to the current times passengers experience. For rental types, we include the two proposed by van Waes et al. (2018), namely one-way and two-way. One-way (OW) rental means the traveller takes the vehicle at one location and leaves it at another. Two-way (TW) means that the travellers must bring the vehicle back to the same location (essentially making a return trip). For parking, van Waes et al. (2018) specifies free-floating (F) or station-based or centralised (C). Here, we add a third option, namely staffed (S) station-based, as we wish to test the difference in travellers' perception to the presence of staff. The full list of attributes and the corresponding levels is presented in Table 1.

*Table 1.Overview of modes, attributes and levels*

|  |  | Walk | BTM | Bicycle | E-Bike | E-Step | E-Moped |
|---|---|---|---|---|---|---|---|
| Walking time [min] |  |  | 2, 6, 10 | 2, 6, 10 | 2, 6, 10 | 2, 6, 10 | 2, 6, 10 |
| Waiting time [min] |  |  | 2, 6, 10 |  |  |  |  |
| Parking type |  |  |  | F, C, S | F, C, S | F, C, S | F, C, S |
| Rental type |  |  |  | OW, TW | OW, TW | OW, TW | OW, TW |
| In-vehicle time [min] | 1km |  | 2, 5, 8 | 3, 6, 9 | 3, 6, 9 | 3, 6, 9 | 2, 5, 8 |
|  | 4km |  | 8, 12, 16 | 12, 16, 20 | 12, 16, 20 | 12, 16, 20 | 8, 12, 16 |
|  | 7km |  | 15, 20, 25 | 25, 30, 35 | 25, 30, 35 | 25, 30, 35 | 15, 20, 25 |
|  | 1km | 10, 15, 20 |  |  |  |  |  |
|  | 4km | 45, 60, 75 |  |  |  |  |  |
| Walking to destination [min] | 7km | 90, 105, 120 | | | | | |
| Cost [€] | 1km | | 0, 1, 2 | 0, 1, 2 | 0, 2, 4 | 0, 2, 4 | 0, 2, 4 |
| | 4km | | 0, 2, 4 | 0, 2, 4 | 0, 4, 8 | 0, 4, 8 | 0, 4, 8 |
| | 7km | | 0, 3, 6 | 0, 3, 6 | 0, 6, 12 | 0, 6, 12 | 0, 6, 12 |

In addition to the different egress distance contexts, we specify two additional variable contexts. Firstly, we vary the duration of the train trip. As Keijer & Rietveld (2000) show, people may be willing to travel longer distances during the egress trip if the main trip is also longer. We define the train travel time contexts as 15min, 45min and 75min, as more than 50% of all trips are between 15min and 45min long, with only 10% being longer than 75min (NS, 2007). Secondly, we vary the purpose of the trip to be either a work/education trip or a leisure trip, where travellers are going to a social activity.

The survey is constructed by using the aforementioned alternatives, attributes and levels in Ngene software (ChoiceMetrics, 2021). For time and cost parameters, we use weak priors (indicating the expected sign only), while for others we specify a zero prior value. We estimate a D-efficient design with blocking, obtaining a total of nine choice sets, split over three blocks. For the contexts, we create an orthogonal design, resulting in 18 combinations, blocked in a way to ensure three context combinations per block. That way, each respondent gets three different context combinations, each of which has three choice tasks, resulting in a total of nine choice tasks per respondent.

We split each choice set into two choice tasks. In the first, respondents are shown only the four SMM alternatives (Figure 1 left). Having made a choice, their selected mode is carried forward into the second task, where it is shown next to the walking and BTM alternatives (Figure 1 right). This reduces the burden on respondents by limiting the amount of information that needs to be processed at once and also provides us with additional insights into what SMM mode respondents would choose if forced to use one and what would they actually use in the end.

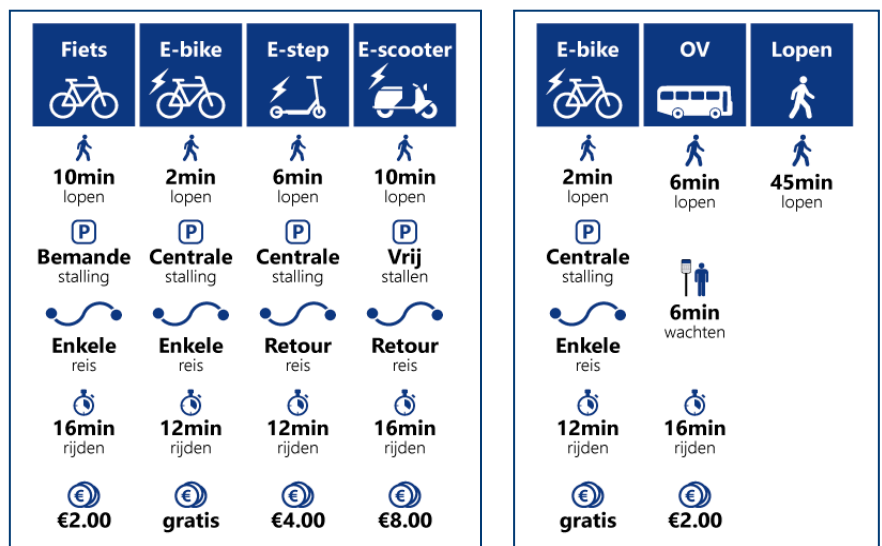

*Figure 1. Example of first (left) and second (right) choice in the experiment*
*(in the example, the respondent chose E-bike in the first choice)*

In addition, we also include attitudinal statements on using SMM in combination with train travel. Adapting the UTAUT2 framework (Venkatesh et al., 2012), we develop a total of 48 statements, relating to technology acceptance constructs as identified in UTAUT2, such as ease-of-use, societal perception, facilitating conditions, etc.. The full list of attitudinal statements can be found in Appendix A.

Finally, we add socio-demographic and travel related question to the survey, to obtain additional insights and to better understand why respondents chose the way they did. Socio-demographics include characteristics such as age, gender, level of completed education, income, household composition etc., whereas travel related questions ask about the usage frequency of different modes and mode preferences for different trip purposes.

## 2.2 Model estimation

The collected choice data is analysed utilising a series pf discrete choice models (DCM), where we assume a utility maximisation approach (McFadden, 1974) by the respondents. The two choice tasks are merged into one when modelling, assuming a single choice among all six options. We estimate a series of multinomial logit (MNL) models with different interaction effects and model formulations to assess their potential impact. Firstly, we test if in-vehicle time is perceived differently for the bicycle, BTM or electric SMM modes. Next, we specify a separate walking time parameter for the walking egress mode and test a logarithmic transformation. For individuals who choose to walk the full egress leg, the marginal contribution of each additional minute may be negligible. Considering the "free" attribute level for cost, we specify it as a dummy to test if there are any additional perceived benefits of a free service. The train trip length and trip purpose contexts are also tested using an interaction effects model specification. The formulation of an interaction effect is presented in Equation 1.

*Equation 1. Interaction effect model specification*

$$(\beta_K + \beta_{AK} \cdot X_{Ac}) \cdot X_{Kc}$$

Where:
- $\beta_K$      parameter for attribute $K$
- $\beta_{AK}$      interaction parameter of attribute $A$ on parameter attribute $\beta_K$
- $X_{Kc}$      attribute level of attribute $K$ in choice task $c$
- $X_{Ac}$      attribute level of attribute $A$ in choice task $c$

To account for respondent heterogeneity, we expand the model into a Latent Class Choice Model (LCCM) (Greene & Hensher, 2003). We opt for the LCCM rather than a mixed logit (ML) model, another common modelling approach accounting for heterogeneity, as it results in clear and distinct population segments. The LCCM allows us to incorporate socio-demographic characteristics into the class membership function of the different classes. The downside of LCCM compared to ML is that they tend to utilise a substantially higher number of parameters and the model estimation is also prone to getting stuck in local maxima. To overcome this issue the model is estimated ten times, with different starting values, in order to minimise the likelihood of the model outcome being a local optimum. The model specification for each of the classes is based on the previously estimated interaction effects and how the individual models performed with respect to the BIC value and LRT test.

To determine the optimal number of classes, we estimate multiple LCCMs with a static membership function (Hess et al., 2008). More classes will likely result in a better fitting model, but that alone may not justify assessing additional classes and model complexity. To that end, we employ a series of numerical and interpretation-related criteria. (1) A lower BIC value is preferred, as it indicates a parsimonious model. (2) Individual classes should be large enough to represent a significant part of the population. We use a rule of thumb of 10% of the sample as the lower bound (Geržinič et al., 2022). (3) All the classes should be interpretable or their behaviour should be understandable. While this last is very subjective, an example can be if all (or most) of the parameters for the class are insignificant, or if the cost parameter is highly positive.

After the optimal number of classes is chosen, socio-demographic information and factor scores are added to the class membership function to better understand the characteristics and attitudes of respondents in the different classes. Parameters analysing the impact of socio-demographic and attitudinal characteristics of travellers are then sequentially removed based on their level of significance, until only significant parameters remain. A parameter is removed if it is insignificant for all classes, while it is kept if it is significant for at least one class

Socio-demographic information is taken directly from the survey, whereas the factors stem from an exploratory factor analysis (EFA) that is carried out on the 48 attitudinal statements from the UTAUT2 framework. A summary can be found in Appendix A with the full survey design and modelling approach reported by Geržinič et al. (2025).

## 2.3 Data collection

The survey was implemented in the Qualtrics survey tool and data collected through two different panels, namely the Dutch Railways own panel (NS Panel) (NS, 2020) and a commercial panel maintained by PanelClix. The NS Panel is used for it's convenience and wide reach among existing train users. PanelClix on the other hand is included to also reach occasional and infrequent train users and to obtain a representative (sub)sample of the Dutch population. Data from both survey panels was collected in summer of 2024, with the NS panel data collected between the 30$^{th}$ of July and 31$^{st}$ of August, and the PanelClix data collected. between the 26$^{th}$ and 30$^{th}$ of August. The former resulted in 2,393 total responses, while the latter leveraged an additional 611 responses.

The data is filtered, removing responses that did not consent to their data being stored and incomplete responses. Next, we check for straightlining behaviour. This is where respondents reply with the same answer to all attitudinal statements, even when this is completely illogical, as some questions are reverse coded. Finally, we remove responses that are deemed too fast to be realistic (Qualtrics, 2024). This leaves us with 1,371 responses from the NS panel and 520 from PanelClix, or a total of 1,891 valid responses to our survey.

An overview of the sample(s) characteristics and the population is presented in Table 2. We can see that overall, the PanelClix subsample is quite well representative of the population as a whole. There is a slight underrepresentation of older individuals (65+), those with a lower (elementary) education. Accordingly, middle-aged individuals (especially 35-49), those with a middle (vocational) education, are overrepresented. Individuals with a driver's license are also somewhat overrepresented in the sample, whereas no clear conclusions can be made for income, due to the fairly high share of those not wishing to disclose their income.

The NS panel sample, on the other hand, is fairly unrepresentative. Although no definitive data exists on this, the NS panel is often used as a proxy for the train-travelling population. As we see in Table 2, the sample tends to be older, with a higher income and very highly educated. Car ownership and consequently driving license ownership are also lower.

*Table 2. Socio-demographic characteristics of the two samples and the population*

|  |  | NS Panel | PanelClix | Population* |
|---|---|---|---|---|
| Gender | Man | 52% | 48% | 50% |
|  | Woman | 48% | 52% | 50% |
| Age | 18-34 | 10% | 26% | 27% |
|  | 35-49 | 23% | 28% | 22% |
|  | 50-64 | 33% | 27% | 25% |
|  | 65+ | 33% | 19% | 25% |
| Household size | One person | 30% | 19% | 19% |

|  | Multiple people | 70% | 81% | 81% |
|---|---|---|---|---|
| Work status | Working | 63% | 69% | 67% |
|  | Not working | 37% | 31% | 33% |
| Education level | Low | 4% | 17% | 29% |
|  | Middle | 20% | 53% | 36% |
|  | High | 75% | 30% | 35% |
| Income | Low | 8% | 18% | 20% |
|  | Middle | 44% | 48% | 45% |
|  | High | 30% | 21% | 35% |
|  | n/a | 18% | 13% | - |
| Driving license | No | 16% | 9% | 20% |
|  | Yes | 84% | 91% | 80% |
| Car ownership | Average | 0.79 | 1.29 | 1.11 |

*\* the population characteristics are based on the >18 population*

With this dual sample, we are able to assess both the preferences of existing users and of the potential new users. All models are estimated on the full sample to leverage the large number of responses we obtained. However, all presentations of latent class characteristics are accompanied by both the sample and population characteristics. What we from here on refer to as population refers to the PanelClix subsample which, as we have shown is quite well representative for the Dutch 18+ population.

## 3 Results

In this section, we present the main findings and outcomes of the survey and model estimation. To give a brief overview of the outcomes and given the dual choice task respondents were faced with (two choice tasks for each trip, see Figure 1), we start with a quick descriptive analysis of the observed choices.

The distribution and migration between modes in the two choice tasks is shown in Figure 2. In general, we see similar proportions from each SMM migrating to walking and PT or sticking with the initially chosen SMM mode, meaning we have a decent representation of all choices and trade-offs between all modes can be observed. Those opting for the bicycle were slightly more likely to shift, particularly to PT, whereas those initially choosing E-moped were least likely to shift, and if they did, they were much less likely to do so to walk.

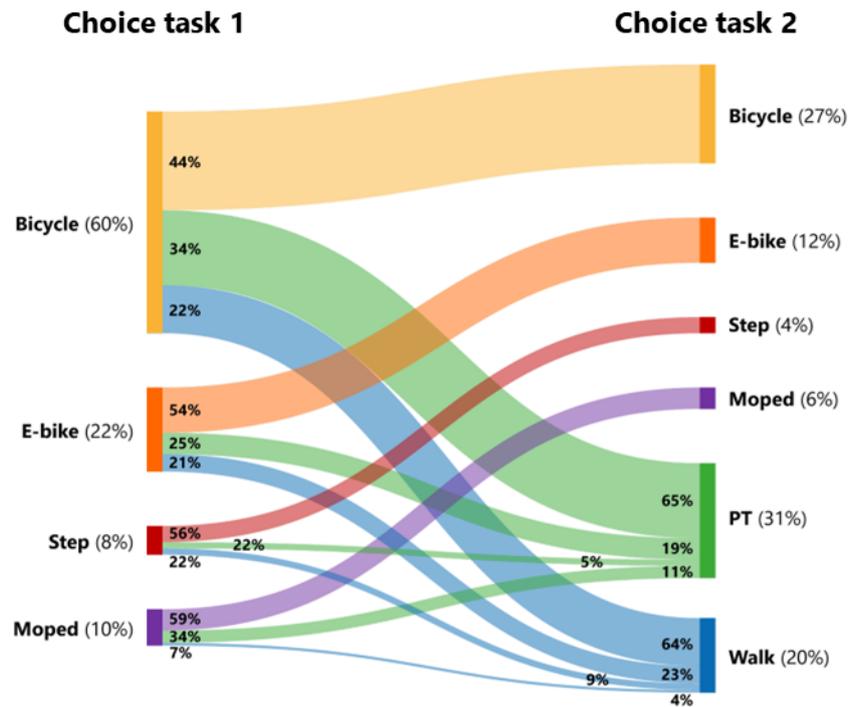

*Figure 2. Graphical representation of choices in the first (left) and second (right) choice task*

## 3.1 Model specification

To estimate all choice models, Biogeme software is utilised (Bierlaire, 2023). To determine the best model specification for the LCCM later on, we start by estimating a baseline model with generic parameters. Thereafter, we test interaction effects and context variables to evaluate their impact on model fit and choice behaviour. The initial model achieves a rho-square of 0.31 (LL = -20981).

Firstly, we assess different in-vehicle time (IVT) specifications by testing if they are perceived differently for different modes, with the bicycle as the baseline. We see that PT IVT tends to be around 10% less negative than cycling, which is to be expected since it requires less physical effort. The parameter is borderline significant with the p-value around 0.05. IVT for electric SMM modes on the other hand is highly insignificant (p=0.58).

Next, we test a variety of walking time specifications, to see if walking within the station is perceived differently than walking all the way to the destination. The interaction parameter for walking to the destination is insignificant, meaning there is no substantial difference in perception of walking time. A logarithmic transformation of walking time also results in an insignificant interaction parameter.

Specifying a dummy cost parameter for when an option is free, results in a strongly significant parameter, with the impact being equal to around €3.00. That means that a free option is perceived equally as receiving a €3.00 discount.

With respect to context, the duration of the preceding train trip has limited to no impact on the choice of egress mode, with the interaction parameters on various attributes (modes, cost, IVT) being highly insignificant. Trip purpose does result in significant parameter estimates, although affecting primarily the ASCs, while seeming not to affect cost perception. The impacts on mode perception make all SMM modes and PT comparatively more attractive than walking when making a leisure/social trip. The relationships between SMM modes however is little affected, as all are shifted by roughly the same scale. Walking time to the destination is also somewhat less negatively perceived which offsets the lower modal preference walking has from the constants.

Based on these outcomes, we define the final model with an interaction effect for PT IVT, free egress trip dummy and walking time interaction for walking all the way to the destination. We initially also include the trip purpose interaction with all the modes, however the model was running into estimation problems and could not be estimated. These interactions are therefore removed from the final model.

Estimating the series of LCCMs with static class membership functions, we proceed with a 3-class model. The 4- and 5-class models had lower BIC values, however we reject the 5-class model because one of the classes is too small (<10%), while the 4-class model is rejected because one of the classes has very limited interpretability of its characteristics (almost all highly insignificant).

## 3.2 Final model outcomes

The final latent class model, including model fit, taste parameters and class allocation parameters is presented in Table 3. In the following paragraphs, the three classes are presented in more detail. The result presentation is done by topic, meaning that for each topic, the results for all three classes are presented, discussed and compared. As stated in Section 2.3, our sample is skewed, however we do have a representative subsample, many statistics are also calculated only for the representative sample, which is from here on referred to as "population".

To ease the analysis, each of the three classes is given a name, based on their preferred travel behaviour characteristics and their attitude towards SMM:

- Class 1: Multimodal sharing enthusiasts
- Class 2: Sharing hesitant cyclists
- Class 3: Sharing-averse public transport (PT) users

*Table 3. Final model outcomes and parameter estimates*

| | **Model fit** | | | | | |
|---|---|---|---|---|---|---|
| Parameters | 81 | | | | | |
| Null LL | -30,493.95 | | | | | |
| Final LL | -17,605.39 | | | | | |
| Rho-square | 0.4227 | | | | | |
| BIC | 35,999.89 | | | | | |
| | **Taste parameters** | | | | | |
| | **Class 1** *Multimodal sharing enthusiasts* | | **Class 2** *Sharing hesitant cyclists* | | **Class 3** *Sharing-averse PT users* | |
| Class size [in sample] | 43% | | 24% | | 33% | |
| Class size [in population] | 58% | | 16% | | 27% | |
| | Est | t-val | Est | t-val | Est | t-val |
| *Constants* | | | | | | |
| Walk | baseline | | baseline | | baseline | |
| Public transport (PT) | -2.830 | -11.00 ** | -4.140 | -11.40 ** | -0.771 | -3.69 ** |
| Bicycle | -2.760 | -13.70 ** | -1.710 | -6.28 ** | -2.720 | -15.10 ** |
| E-Bike | -2.420 | -11.10 ** | -4.480 | -14.00 ** | -3.190 | -15.00 ** |
| E-Scooter | -3.570 | -15.80 ** | -6.800 | -15.50 ** | -4.560 | -18.50 ** |
| E-Moped | -3.380 | -14.70 ** | -7.930 | -13.20 ** | -5.020 | -18.30 ** |
| *Taste parameters* | | | | | | |
| In-vehicle time | -0.052 | -9.80 ** | -0.044 | -4.49 ** | -0.046 | -6.44 ** |
| In-vehicle time [PT interaction] | 0.025 | 3.95 ** | 0.052 | 4.23 ** | 0.011 | 1.51 |
| Waiting time | -0.077 | -7.02 ** | -0.068 | -3.44 ** | -0.039 | -4.47 ** |
| Walking time | -0.139 | -21.10 ** | -0.055 | -3.91 ** | -0.083 | -9.09 ** |
| Walking only [interaction] | -0.009 | -0.74 | -0.072 | -4.38 ** | 0.016 | 1.64 |
| Cost | -0.229 | -11.00 ** | -0.211 | -5.05 ** | -0.199 | -7.38 ** |
| Free trip | 0.963 | 9.73 ** | 0.383 | 2.33 * | 0.111 | 1.04 |
| Free floating parking | -0.042 | -0.79 | -0.085 | -0.64 | 0.312 | 3.08 ** |
| Staffed facility | -0.044 | -0.86 | -0.017 | -0.18 | 0.279 | 2.92 ** |
| Two-way rental | 0.259 | 5.76 ** | 0.184 | 2.22 * | -0.020 | -0.23 |
| **Class membership parameters** | | | | | | |
| Constant | | | 0.477 | 0.93 | 0.268 | 0.53 |
| Female gender | | | -0.453 | -2.92 ** | 0.229 | 1.57 |
| Age 18-34 | | | -0.649 | -2.16 * | -0.053 | -0.21 |
| Age 50-64 | | | 0.509 | 2.50 * | 0.235 | 1.17 |
| Age 65+ | | | 0.564 | 2.27 * | 0.154 | 0.65 |
| Retired | | | 0.307 | 1.22 | 0.723 | 3.09 ** |
| Smartphone ownership | | | -1.350 | -2.93 ** | -1.310 | -2.87 ** |
| Experience with shared bicycle | | Baseline | 0.841 | 4.47 ** | -0.144 | -0.84 |
| Off-peak subscription | | | 0.223 | 1.39 | 0.383 | 2.41 ** |
| Weekly BTM use | | | -0.030 | -0.14 | 0.884 | 4.98 ** |
| Weekly car use | | | -0.636 | -4.09 ** | -0.633 | -4.14 ** |
| Station access by BTM | | | -0.216 | -0.85 | 0.674 | 3.23 ** |
| Station access by car | | | -0.463 | -1.62 | 0.690 | 2.82 ** |
| Station access walking | | | -0.115 | -0.61 | 0.769 | 4.16 ** |
| F1 Intention to use SMM | | | 0.120 | 1.09 | -0.455 | -4.61 ** |
| F4 SMM has a good societal image | | | -0.180 | -2.09 * | -0.358 | -4.20 ** |
| F6 Using PT is healthy | | | 0.192 | 2.16 * | 0.482 | 5.31 ** |
| F7 Moped is fun and safe | | | -0.973 | -8.63 ** | -0.971 | -8.94 ** |

*\*\* p ≤ 0.01, \* p ≤ 0.05*

**Modal preferences**

Starting with modal preferences based on the ASCs, we can se that all three classes have the strongest preference for walking, as all the parameters are negative. This is to be expected, since if travel time is equal, the vast majority of individuals are likely to prefer to walk. For longer egress distances, walking quickly becomes too slow and other modes become comparatively more attractive. Is it therefore interesting to compare the modes (PT and SMM) amongst each other, as they primarily all compete at distances beyond what is comfortable to walk. Figure 3 shows the WtP for travelling compared to PT, all other attributes being equal. A positive value indicates that mode is in preferred over PT, while a negative value means that PT is preferred compared to the respective mode. Notably, the multimodal class has the smallest overall preferences, with WtP values below €5.00. That means they do not have strong preferences and will primarily base their decisions on the characteristics of each mode, choosing the one that has the best time-cost ratio. Interestingly, they are also the only ones where the E-bike is preferred over a regular bicycle. The other two classes have stronger travel mode preferences, with their names indicating which is their mode of choice (bicycle or public transport). Also interesting is that the step and moped are last for all three classes.

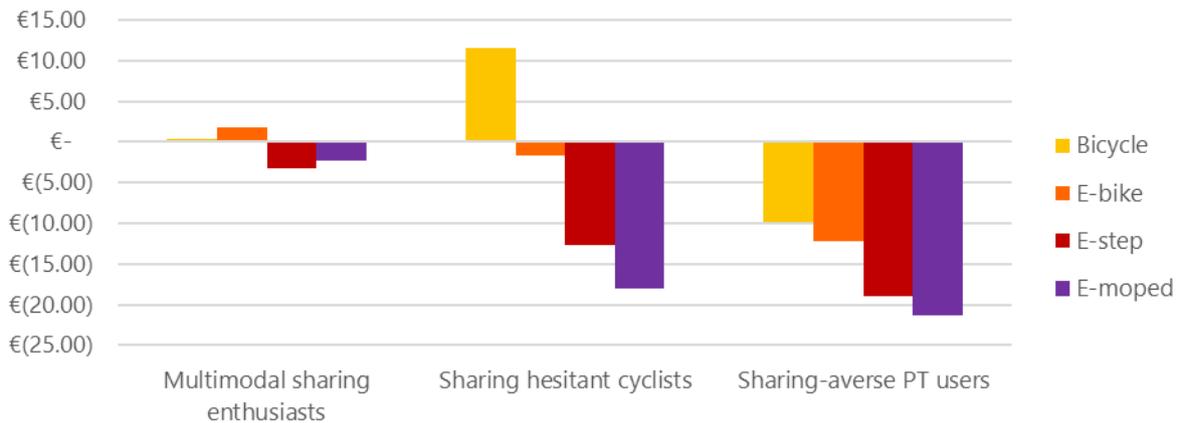

*Figure 3. Willingness-to-Pay (€) to travel by a different mode, instead of with public transport (ceteris paribus)*

**Travel time components**

In the survey, we showed the respondents many different travel time components, with the WtP of all presented in Figure 4. Interestingly, all three classes have a very similar value of IVT at ~13€/h. Where they differ though is in their perception of other parameters. Looking at the PT IVT, the multimodal group see it only half as negative as the IVT on SMM, possibly indicating they see the benefit of using that travel time productively or the fact they do not have to exert as much physical effort when taking PT. The cyclists seem to not mind the travel time at all with PT, as it completely cancels out the original parameter. The big difference in ASC still ensures most members of this class will choose the bicycle, but when compared to other modes, PT seems to be a strong second. It also indicates that travelling by PT for this group carries an constant barrier, which is unrelated to the duration of the trip (at least for the tested range of attribute levels). For the Sharing-averse PT users, the situation is inverse, with IVT perception not differing between PT and other modes. Their strong initial preference for PT however ensures that this will still be their mode of choice.

Walking is perceived as the most negative for all classes, with a factor of 2-3x that of IVT, which is also in line with expectations. Walking perception only differs for the second class and curiously, walking within the station is perceived less negatively than the whole route to the destination. Waiting time perception is on the low side, at around 1.5x the IVT perception and even lower than the IVT for the PT users.

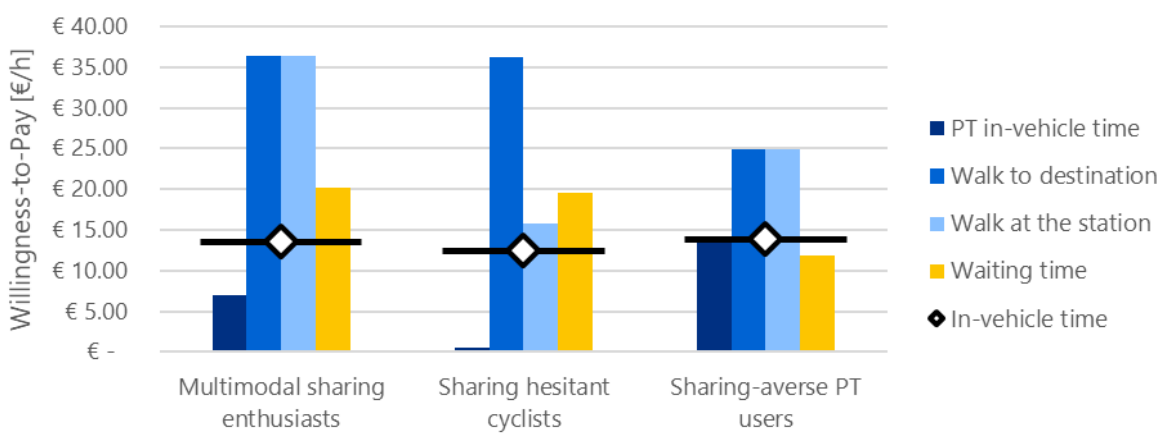

*Figure 4. Valuation of travel time components by different classes*

**SMM design characteristics and free egress mode**

As we see from Table 3, the "free" dummy parameter and the two-way trip are significant for the first two classes, while the presence of staff and free-floating parking seems to be relevant for the third class. To better understand the trade-off values for these characteristics, Table 4 shows how much walking time and cost are the different classes willing to trade. Starting with "free", we see that the first group really see this as a big benefit, equal to over €4 of discount. Both the multimodal and cycling classes would also be willing to walk 7min further to access a service which is free, which is a substantial distance. Interestingly, the same two groups also see added value in a two-way rental scheme, willing to pay €1 extra or walk 2-3min farther to use such a service. On the other hand, only the PT users see the presence of staff beneficial to such an extent it influences their choices. We can see they would be willing to pay €1.41 extra when renting SMM if staff was present, or walk an additional 3min. As we see later, this group is more anxious and less experienced with shared services and thus preferring the presence of staff is logical. More interestingly, this group seems to prefer a free-floating parking arrangement, compared to all vehicles being available in one centralised location. This goes against our expectations, and it may also mean they did not fully understand the concept, potentially because they are less experiences SMM users.

*Table 4. Trade-off values of walking time and cost in return for different SMM design characteristics or a free service*

|  |  | Multimodal sharing enthusiasts | Sharing hesitant cyclists | Sharing-averse PT users |
|---|---|---|---|---|
| Free | Cost | € 4.21 | € 1.81 | *€ 0.55* |
|  | Walking time | 7min | 7min | *1min* |
| Two-way trip | Cost | € 1.13 | € 0.87 | *€ -0.10* |
| *(baseline: one-way)* | Walking time | 2min | 3min | *-0min* |
| Staffed facility | Cost | *€ -0.19* | *€ -0.09* | € 1.41 |
| *(baseline: non-staffed)* | Walking time | *-0min* | *-0min* | 3min |
| Free-floating parking | Cost | *€ -0.19* | *€ -0.40* | € 1.57 |
| *(baseline: central parking)* | Walking time | *-0min* | *-2min* | 4min |

**Attitudinal statements**

In addition to behavioural data, we also collected attitudinal statements from respondents. We then performed an exploratory factor analysis (EFA) to reduce the 48 items into 8 factors. The EFA is performed on the full sample and the individual factor scores then recomputed for each of the three classes and also for the representative subsample to get an idea how the population scores. The full EFA procedure is outlined in more detail in Appendix B. In Figure 5, the population score is taken as the baseline, with the deviations of the full sample and individual classes. From this analysis, all three classes obtain the part of their name referring to their attitudes towards SMM. The first class generally has a strong intention of using SMM, they think it's easy to use, are confident using apps and smartphones (F8), and compared to other classes and the sample average, they perceive it as fun and safe. They are somewhat less confident about the availability of SMM vehicles, although the most of all classes and they also believe SMM has a decent societal image. The sharing hesitant group shows potential and openness to it, scoring highly on usage intent, societal image of SMM and climate awareness, but has some reservations towards using it, primarily due to the safety concerns, ease-of-use and use of technology. Finally, the sharing-averse group does not see themselves using it at all, are least confident about availability, find it the most difficult to use out of any group, the most dangerous and are also the least confident in being able to use it or in using smartphones and apps.

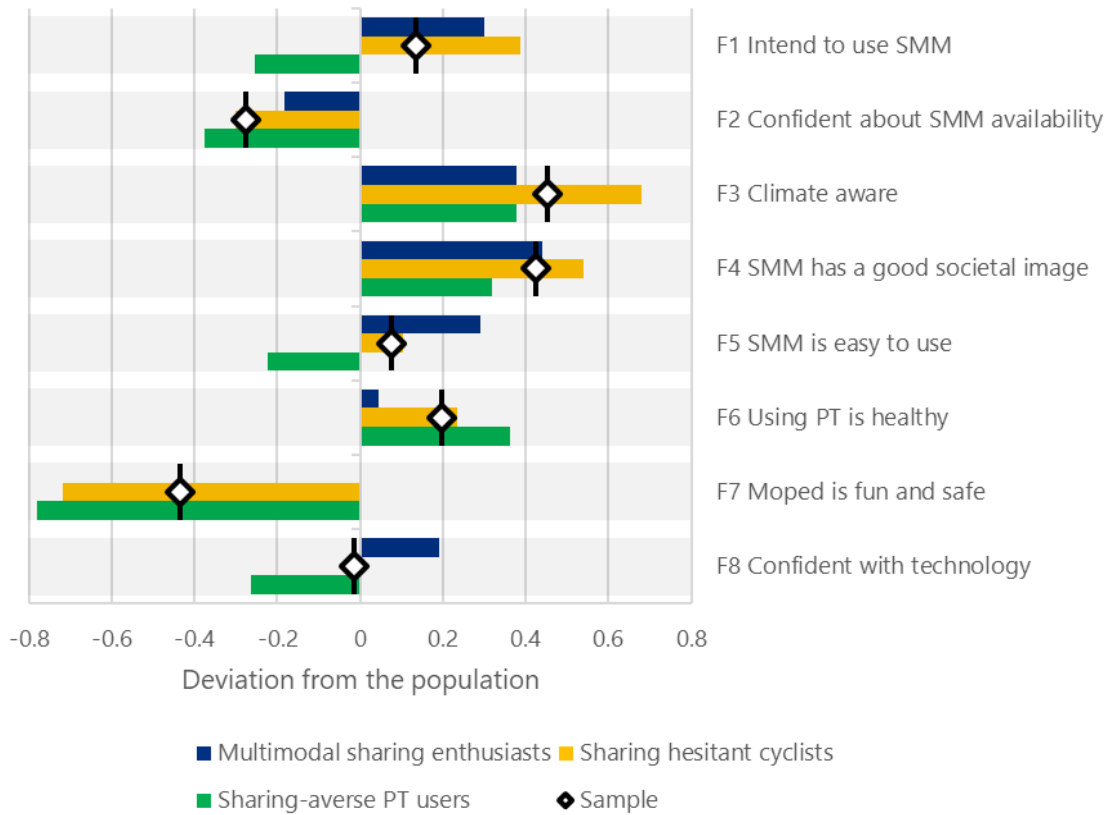

Figure 5. EFA scores of the classes compared to the sample and population

**Current travel behaviour (matches their stated behaviour)**

Next, we move to the respondents' current travel behaviour. The modes they use on a weekly basis (Figure 6) match their uncovered preferences in the stated choice experiment, with the multimodal group having the highest shares in combination categories (bicycle+PT, bicycle+car, PT+car). The cycling group has high shares in all categories where cycling is present, and the same can be said for the PT group about their preferred mode. It is also interesting to see how all three groups differ from the distribution of the population, which highlights the substantial difference between regular and occasional/non-train users.

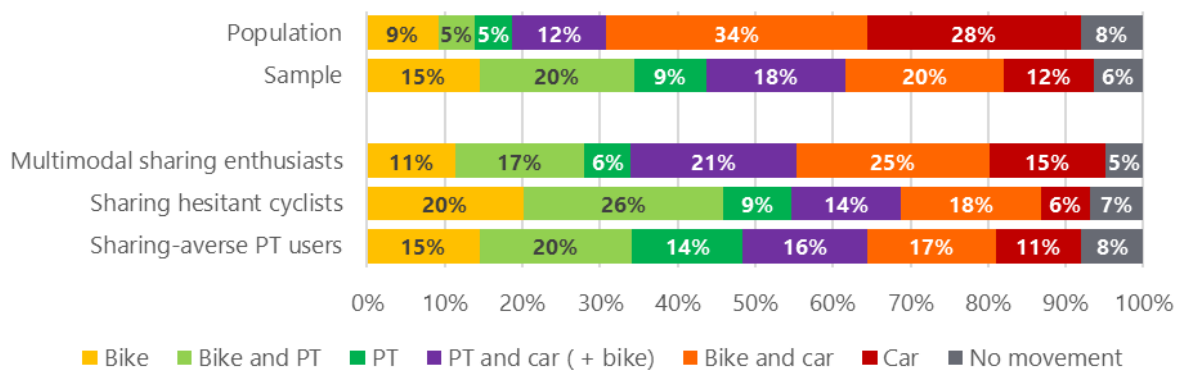

Figure 6. Modes used on a weekly basis by the different classes, compared to the sample and population averages

The names also correspond when analysing their preferred mode on the access-/home-side of the trip (Figure 7), with the multimodal group having a distribution most similar to the sample overall, the cycling group using that mode in >60% of instances, and the PT group walking and using PT more than most.

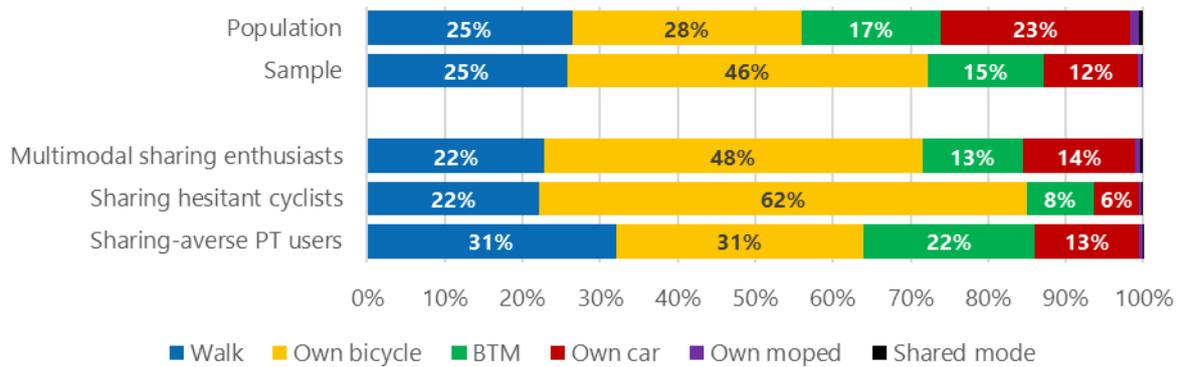

*Figure 7. Train station access (home-end) among the classes, the full sample and the population*

Finally, we also compare the respondents' experience and familiarity with different shared mobility services, as shown in Figure 8. Bicycle sharing is clearly well-known and most respondents have experience using it. This primarily refers to the Dutch Railways' own bicycle sharing service OV fiets, which is widely available around the country and well integrated into the train ticketing system. This also explain why the sharing hesitant group is the most familiar here, as although they have some concerns about SMM, once a service is widely available, easy to use and well integrated, they are open to using it. The same could be said about the fact that they are the most experienced car sharing users. The sharing enthusiasts on the other hand top the experience list on modes which are less prevalent in the Netherlands. Consistent with their names, the sharing-averse class has least experience with all shared modes. In most cases, they are below both the sample and population averages also, except for bicycle sharing. This also indicates just how well integrated this service has become, as even sharing-averse train users use it more than the population on average.

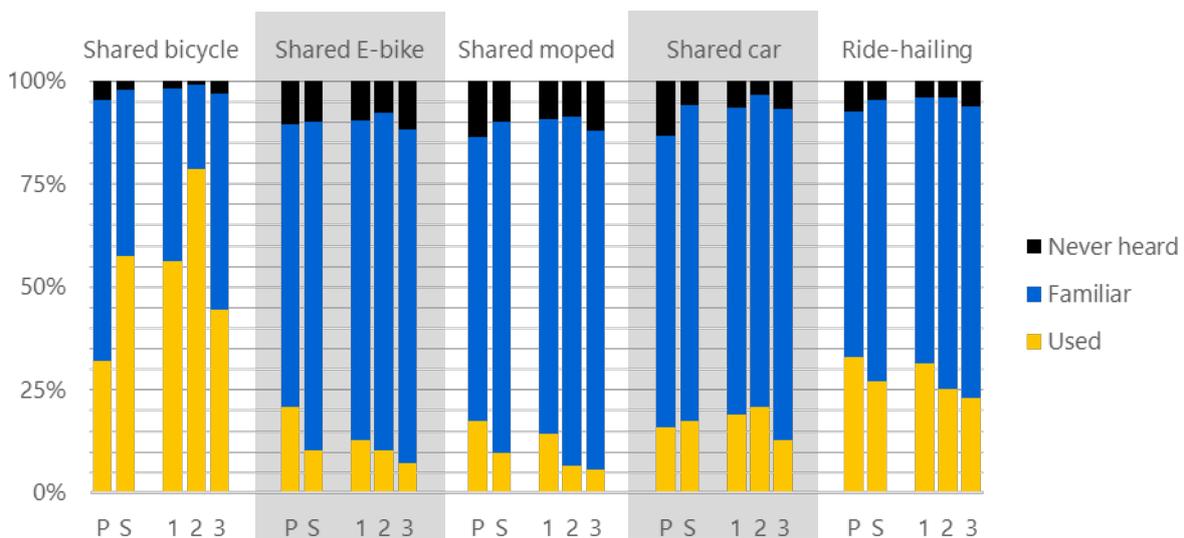

*Figure 8. Experience with different shared mobility services*
*(P = Population, S = Sample, 1 = Class 1: Multimodal sharing enthusiasts,*
*2 = Class 2: Sharing hesitant cyclists, 3= Class 3: Sharing-averse PT users)*

**Socio-demographic characteristics**

Lastly, we look into the socio-demographics of the classes and compare them to the sample and population, which is presented in Table 5. Starting with The multimodal sharing enthusiasts, they are predominantly characterised by being younger than average and the youngest of the three groups, in all groups below 50 years old. Given this age, it is no surprise they live in multi-person households (with a partner and children usually) and have the highest share of individuals working. They also have the highest car ownership of any class, although still somewhat below the population average. They are quite highly educated, with a high income.

The two other classes are both somewhat older and have an above average share of retired individuals. Members of the sharing sceptic cycling class tend to be the highest educated and have the highest income. At the same time, they have the lowest car ownership and live in multi-person households, primarily with a partner only (kids already moved out).

The sharing-averse PT users on the other hand are the most likely to live alone, have the lowest level of education and lowest income of any class. They are also least likely to have a driving license and a slightly above average share of the class's members are female.

*Table 5. Socio-demographic characteristics of the sample, population and classes*

|  |  | Sample | Population* | Class 1 | Class 2 | Class 3 |
|---|---|---|---|---|---|---|
| Gender | Man | 51% | 50% | 53% | 58% | 44% |
|  | Woman | 49% | 50% | 47% | 42% | 56% |
| Age | 18-34 | 14% | 27% | 20% | 7% | 12% |
|  | 35-49 | 24% | 22% | 30% | 20% | 20% |
|  | 50-64 | 31% | 25% | 31% | 36% | 28% |
|  | 65+ | 31% | 25% | 20% | 37% | 40% |
| Household size | One person | 24% | 19% | 20% | 23% | 31% |
|  | Multiple people | 76% | 81% | 80% | 77% | 69% |
| Work status | Working | 52% | 67% | 60% | 49% | 45% |
|  | Not working | 48% | 33% | 40% | 51% | 55% |
| Education level | Low | 8% | 29% | 8% | 5% | 10% |
|  | Middle | 30% | 36% | 33% | 22% | 31% |
|  | High | 63% | 35% | 60% | 73% | 59% |
| Income | Low | 11% | 20% | 10% | 9% | 13% |
|  | Middle | 45% | 45% | 44% | 45% | 46% |
|  | High | 28% | 35% | 30% | 31% | 22% |
|  | n/a | 17% |  | 15% | 15% | 19% |
| Driving license | No | 14% | 20% | 11% | 15% | 18% |
|  | Yes | 86% | 80% | 89% | 85% | 82% |
| Car ownership | Average | 0.93 | 1.11 | 1.06 | 0.82 | 0.83 |

*\* here, the "Population" category is not the representative subsample characteristics, but true population characteristics*

# 4 Policy implications

To better understand how travel behaviour is affected and what the impact of introducing of new SMM services is, we perform a series of sensitivity analyses and modal split calculations. As there are many different types of train stations, with different egress travel availabilities and attribute levels, we use the Dutch train station classification by de Wit et al. (2024). Specifically, we choose to analyse *Central station in other larger towns* (Type B according to de Wit et al. (2024)). We focus on these medium-sized towns (~100.000 inhabitants) as they tend to have substantial attraction potential (many people travelling there) while having a public transport service that is of lower quality than in the largest cities. The combination of these two factors means there is a high potential for SMM services. Although other station types, like *Local train stations in villages and outer areas* (Type H as classified by de Wit et al.

(2024)) would likely benefit more from the introduction of SMM, their low passenger numbers mean that the overall added value would likely be smaller.

For the analysis, we take typical characteristics for Central stations in other large towns, as presented in Table 6. In order to calculate modal shares for varying distances, we use an approximation for determining travel time and cost. For time, we specify the average speed of each mode and derive the travel time from that. We apply a similar approach for cost, based on current pricing strategies, where PT has an initial fee and a distance-based rate, the shared bicycle and e-bike have a flat rate only, while the e-moped has a starting fee plus a time-based fee. As e-steps are currently not available in the Netherlands, we use the same characteristics for it as for the e-moped. Waiting time for PT is based on average headways of 15min. Walking time for all modes assumes it takes approximately 2min to walk from the platform to the vehicle/bus stop, and for PT an additional 5min from the bus stop to the final destination. Rental and parking types are also based on current SMM service characteristics. Finally, as we wish to test how the quality of PT affects SMM, we also test a "Lower quality PT" scenario, where the walking and waiting times are increased. Specifically, waiting is increased to 15min, mimicking a 30min headway, and walking to 12min, taking 2min from the platform to the bus and 10min from bus to destination.

*Table 6. Attribute levels used for calculating egress mode utilities*

|   |   | Walk | PT | Bicycle | E-bike | E-step | E-moped |
|---|---|---|---|---|---|---|---|
|   | Average speed *[km/h]* | 4 | 20 | 12 | 15 | 12 | 15 |
|   | Waiting time *[min]* |   | 7.5 *(15)* |   |   |   |   |
|   | Walking time *[min]* |   | 7 *(12)* | 2 | 2 | 2 | 2 |
| Cost | Initial fee *[€]* |   | 1.08 | 2.3 | 6.5 | 1 | 1 |
|   | Tima rate *[€/min]* |   |   |   |   | 0.33 | 0.33 |
|   | Distance rate *[€/km]* |   | 0.18 |   |   |   |   |
|   | Rental type |   |   | Two-way | Two-way | One-way | One-way |
|   | Parking type |   |   | Staffed | Staffed | Free-floating | Free-floating |

*(values in italic red brackets) indicate the "Lower quality PT" scenario*

## 4.1 Attribute sensitivity

In this section, we look at how different classes would react for different trip distances and under different circumstances. Figure 9 shows the evolution of modal split with an increasing egress distance, with the top row showing the initial or higher quality PT scenario while the lower shows the lower quality PT scenario. Firstly, we observe that walking dominates on shorter distances, dropping below 50% somewhere around 1.3km, with Class 3 being the most enthusiastic walkers and Class 2 the least. Secondly, as expected, cycling seems to be most popular for distances of ~5km, losing ground to PT and E-bikes at longer egress distances. Thirdly, the drop in PT quality benefits cycling for shorter trips (up to 10km) and e-bikes for trips over 10km, although both modes see an increased market share as a result. Fourthly, e-steps and e-mopeds seem to be most attractive for trips of ~2km, which is somewhat surprising. This is primarily due to their pricing strategy, which discourages longer trips. While they may be more comfortable and in some cases faster than cycling, the latter is currently priced with a fixed rate, meaning the disutility of longer trips is offset by the proportionally lower price (less per km). When the price also includes a variable component (time or distance based), this seems to have a negative impact as then both time and cost increase, meaning they are becoming comparatively less attractive to other SMM modes. PT suffers from this to a lesser extent due to its higher travel speed and lower penalty for IVT of some travellers. Finally, the modal preferences of the different classes are very noticeable. Class 1 exhibits a distributed modal split and highly sensitive behaviour when circumstances change. Classes 2 and 3 on the other hand demonstrate just how strong their preferences for cycling and PT respectively are and that their sensitivity to PT-related attributes is lower.

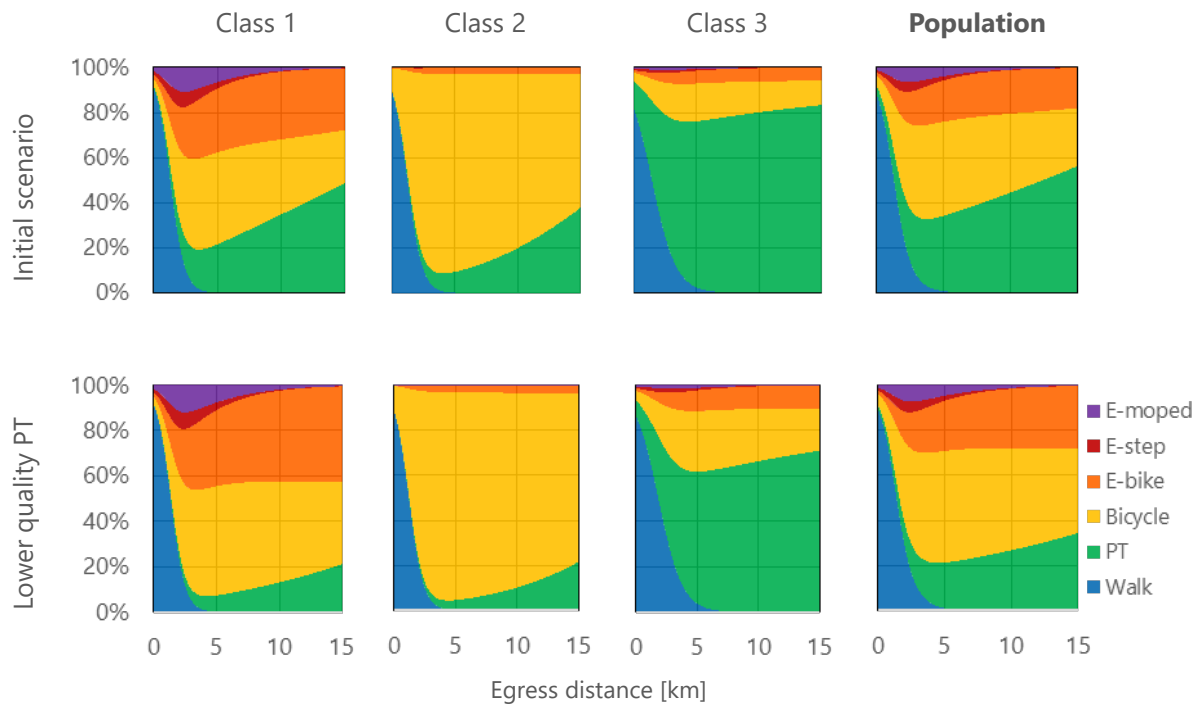

*Figure 9. Modal split as a function of egress distance*

## 4.2  Introducing new SMM services

In this section, we analyse the effects associated with the introduction of new SMM, focusing on the modal shifts and added value of these services for consumers. Starting with modal shift, we observe the migrations between modes in Figure 10. The figure shows the shifting of travellers if new modes were introduced consecutively, starting with the shared bicycle, e-bike and then e-moped. At an egress distance of 2km, the average for a *Central station in other larger towns* (de Wit et al., 2024), the majority of individuals would opt to walk to their final destination if no SMM is available. Adding a shared bicycle would drastically change this, making it the most attractive option (43%), with the majority shifting from walking, halving it's market share, whereas PT would only lose a third. Afterwards, the attractiveness of additional SMM services is lower. Adding E-bike attracts around 14%, mainly from those walking or cycling. Also including E-moped would attract around 7%, again mainly from other SMM and walking. In all the instances, PT contributes the least passengers and also proportionally loses the least. Walking and cycling tend to lose equal shares to both e-bike and e-moped, while when adding e-moped, e-bike loses more of its market share: 10%, compared to 7% for walking or cycling and 5% for PT.

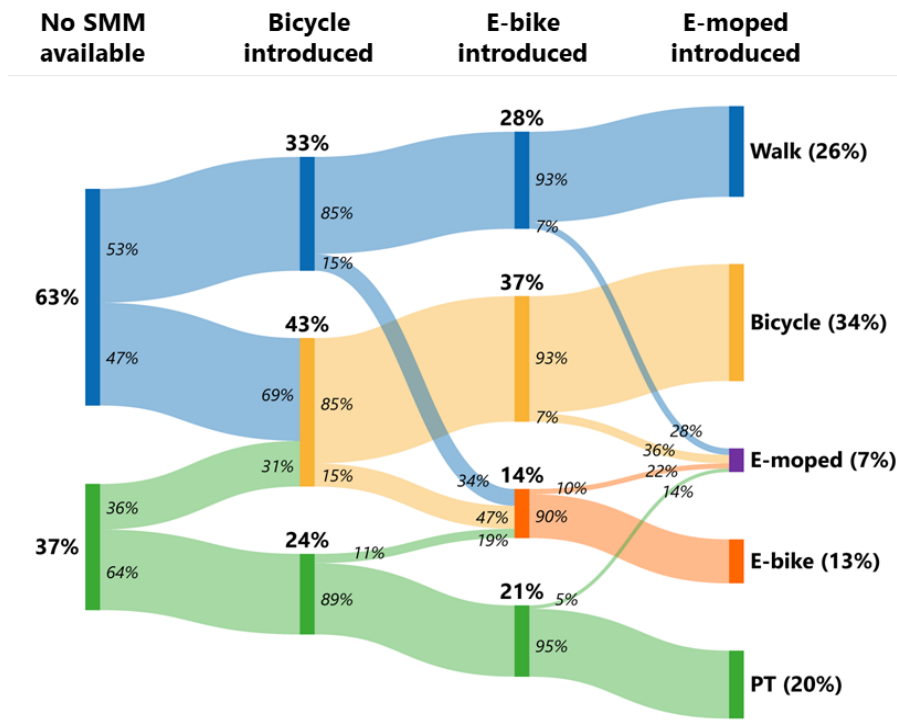

*Figure 10. Modal shift when new SMM are introduced (assuming the average egress distance of 2km)*

## 4.3 Policy recommendations

Based on these findings, we formulate several recommendations on which modes to focus on in different circumstances, how to approach service design and how to manage larger portfolios of SMM modes.

Firstly, we see that walking is the preferred mode on shorter distances from the station, up to ~2km. As highlighted by previous research already, this underscores the importance of transit-oriented development. For policymakers, this reiterates the importance of policies relating to densification of land-use around train stations to enable as many people as possible to work, reside or have their other activities within walking distance from a train station. Policies should also focus on making the areas around train stations attractive and safe to walk in.

For longer distances, the preferred mode is less unanimous and more dependent on the specific circumstances of the train station, trip characteristics and the individual's preferences. In general, PT and the bicycle tend to be preferred for distances up to ~10km, from where PT and E-bike tend to dominate. As we can see in Figure 9, the quality of PT can substantial impact the modal split, with the example showing a significant drop in PT market share when the quality was reduced (longer wait and walk times). Providing high-quality PT (with dense and frequent network), can be expensive and can therefore often not be economically justified due to lower demand and ridership.

We therefore recommend policymakers and operators to focus on local PT at stations where land-use density and thus demand are sufficiently high, meaning that providing high quality service can be justified. This is also advantageous from the SMM perspective, as high-demand areas will require a larger fleet of SMM vehicles to satisfy the high demand.

At stations where high quality PT cannot be justified, operators and policymakers should focus primarily on providing SMM services, since these are more attractive to the majority of travellers than limited PT service. Depending on the distances egress from the station, bicycles and e-bikes should be provided as necessary.

Most stations are likely a combination of the two, since certain corridors will have sufficient density to justify high quality PT while other directions may not. That means that both PT and SMM should be provided to cater to travellers headed to different destinations. SMM will likely still be used for all trips from the stations, but travellers heading in the direction where PT is attractive will be drawn to it more, meaning most SMM trips are likely to serve to trips to lower demand areas.

For e-steps and e-mopeds, in the current pricing regime, they primarily fill a niche market between walking and cycling, at distances of ~2-3km. Should these want to be encouraged further, a different pricing scheme is likely needed. Although given the low interest in these alternatives by all three user groups, policymakers should not put too much focus on these being the core SMM modes.

Pricing can have a substantial impact on the attractiveness of travel alternatives and can thus be used to steer behaviour. We see that alternatives tend to perform better when priced with a flat fare only (bicycle and e-bike) rather than proportional to time/distance travelled (e-moped and e-step). This latter pricing approach is only justified if the alternative offers substantial benefits to offset the higher overall travel cost. This offset can be in terms of travel time and comfort, as is the case with (higher quality) public transport. Flat fares are also beneficial as they discourage use of SMM for short trips and instead steer travellers to walking instead. Policymakers should therefore encourage flat fares as these can achieve high attractiveness for longer trips by making them comparatively cheaper, while at the same stimulating travellers to walk for shorter distances. For operators, proportional pricing may be more advantageous to also attract shorter trips, but then a more careful calculation of the specific combination of starting price and time/distance-based fare is needed.

In the current Dutch situation (as modelled here), the shared bicycle tends to be the most preferred SMM mode, partly due to its low cost and widespread acceptance and familiarity. Given such high demand, operators and policymakers should put effort into guaranteeing that sufficient vehicles are available, meeting the mobility needs of most travellers. This is particularly important at stations where the majority of egress travellers rely on SMM.

When introducing new modes operators and policymakers should focus on providing options which are cheaper for them to provide. Bicycles tend to be cheaper than electric SMM to procure, meaning they can, in turn, provide it to the travellers for less. The low cost of renting a shared bicycle is one of the primary reasons why it is the most popular mode for short egress distances.

Adding additional SMM modes at stations where some already exist requires a more fine-tuned service design. Electric modes (particularly the E-bike) can complement regular bicycles very well by providing a lower physical effort alternative, especially over longer distances. E-mopeds and e-steps could likely achieve similar, although they are likely to attract fewer travellers due to them being less attractive as modes. The issue with adding new SMM modes is that they tend to compete with each other. We see in Figure 10 that new SMM alternatives are more likely to attract users from existing SMM modes. This can be overcome to a certain extent through well-defined pricing strategies. We see from the example that, despite both bicycles and e-bikes being priced with a flat rate, the latter becomes more attractive as the travel time savings add up, making the higher price worthwhile. Alternatively, operators and policymakers could adopt a distance-dependent pricing approach to better stimulate the use of specific modes as desired.

# 5  Conclusion

In this research, we analyse passengers' perception and potential of shared micromobility (SMM) services (bicycle, e-bike, e-step and e-moped) as a solution for the last mile of train trips. Our results show three distinct user groups with respect to mode choice behaviour and attitudes towards SMM on the activity side of a train trip. *Multimodal sharing enthusiasts* (58% of the population) tend to make decisions based primarily on the trip characteristics and are open to switching modes if the cost-time-comfort trade-offs are better in their eyes. They are positive about SMM and generally more open to using it than the other two groups. *Sharing hesitant cyclists* (16% of the population) have a strong preference for the bicycle and would rarely choose to travel by public transport (PT). They are still positive about SMM, have quite some experience with it (particularly shared bicycles) and do intent to use it, but have some reservations about it, primarily relating to safety and ease-of-use. *Sharing-averse PT users* (27% of the population) are mostly likely to travel by PT, having a strong aversion to all SMM. In that regard, they do not intend to use it either as they find it difficult, dangerous and having a bad societal perception. Their lack of experience with it and the lack of confidence using smartphones is also reflected in their willingness-to-pay for the presence of staff when potentially renting an SMM vehicle.

For policymakers, our findings suggest that walking is the most preferred egress mode and the dominant choice for short distance trips (<2km), meaning that land-use densification policies like transit oriented development should be encouraged. For longer egress distances, local PT like buses and trams should be encouraged where economically feasible and justifiable (sufficient ridership), while SMM in all other cases. Existing SMM alternatives (bicycle and e-bike) tend to be preferred to e-steps and e-mopeds. Finally, we observe a strong influence of pricing schemes on mode choice, with flat rates performing best from a societal perspective, as they encourage walking over short distances.

Our findings are broadly in line with other research on the topics of shared mobility, SMM and Mobility-as-a-service (MaaS). Most studies in the Dutch context (Alonso-González et al., 2020; Geržinič et al., 2022, 2023; Montes et al., 2023; Winter et al., 2020) report that a large part of the population (usually the largest group stemming from a segmentation analysis) tend to be open to new forms of mobility. They see the benefits of it, feel themselves capable of using it and being digitally savvy. What somewhat differs in our study is the sharing-averse group, which seems to be highly pro-PT, whereas the above referenced studies tend to report such groups as highly car-dependent. Alternatively, the high pro-PT groups found in other studies tend to be highly sharing motivated, especially in the results of Alonso-González et al. (2020) and Geržinič et al. (2022). In this study however, the cycling group was also quite positive about many aspects of shared mobility. This group (*Sharing hesitant cyclists*) is also more similar to the enthusiastic sharing groups of other studies based on socio-demographic characteristics, specifically having a lower car ownership, on average older and more likely to be female.

The differences between studies may be due to our sample being biased towards train users. While still having a representative subsample, the full sample is skewed towards more frequent train users, meaning that all results are likely also skewed in that direction. This is noticeable through the overall lower car ownership and car use, as none of the classes is highly car-dependent, something that appears throughout all other studies. Future studies should therefore also capture car users who rarely or never travel by train. A study on this topic could also overcome the second assumption of our study, namely that we assumed taking the train on the main leg of the trip is given. The combination of service quality on the full trip chain access-main-egress is often what influences the choice to take the train or rather a private car or not to travel at all. We did not give respondents the opportunity to opt-out of the train trip as we were primarily interested in the activity-end of the trip, but the quality and availability of activity-end travel can make or break the full trip chain. Future research should therefore investigate further how important this is to attracting more people onto the train and out of the car.

## Acknowledgments

We would like to thank the Dutch railways for their support in carrying out this research. We also want to thank all the respondents of the panels who invested their valuable time and energy in the survey which helped us to carry out this research.

# Appendices

## A. Attitudinal statements

Attitudinal statements are developed based on the constructs defined in the UTAUT2 framework (Venkatesh et al., 2012), which is a frequently used and cited technology use and acceptance model. We adjust the constructs and develop 3-6 statements for each of the constructs. The full list of constructs is provided below:

1. Behavioural intention
   1.1. I intend to use shared micromobility services when travelling by train
   1.2. I intend to use shared micromobility when going to work or education.
   1.3. I intend to use shared micromobility when visiting friends/family.
   1.4. I would travel by train more if I had more shared mobility options to get to/from the station.
   1.5. I would travel by train more if I had more public transit (e.g. bus/tram/metro) options to get to/from the station.

2. Performance expectancy
   2.1. I believe that using shared micromobility will save me time when travelling.
   2.2. I believe that using shared micromobility will make my travel less efficient than it is now.
   2.3. I believe that using shared micromobility will save me money.

3. Effort expectancy
   3.1. I expect it will be easy for me to learn how to use a shared (electric) bicycle.
   3.2. I expect it will be easy for me to learn how to use a shared electric moped.
   3.3. I believe I will not have problems unlocking shared (electric) bicycles on my own.
   3.4. I believe I will not have problems unlocking shared electric mopeds on my own.
   3.5. I think it is easier to use shared micromobility if the vehicles are all parked together in the same location.
   3.6. I think it is difficult to find information on how to use shared micromobility (sign-up, create an account, unlock a the vehicle,…).

4. Social influence
   4.1. I can see myself using shared micromobility.
   4.2. My public image (how people see me) is important to me.
   4.3. My friends would think less of me if I used shared micromobility.
   4.4. My family would think less of me if I used shared micromobility.
   4.5. I believe it is societally responsible to use shared micromobility.

5. Facilitating conditions
   5.1. I have a smartphone. (move to socio-demographics)
   5.2. I know how to use smartphone applications.
   5.3. I have smartphone applications for (one or more) travel companies on my smartphone.
   5.4. I do not mind having multiple different applications for different travel companies on my smartphone.
   5.5. I would prefer unlocking shared micromobility vehicles using a card (e.g. OV chipkaart) and not a smartphone application.
   5.6. I do not mind making payments through smartphone applications.

6. Hedonic motivation
   6.1. It is fun to use a shared (electric) bicycle.
   6.2. It is fun to use a shared electric moped.
   6.3. I can enjoy my surroundings when I travel by (electric) bicycle.
   6.4. I can enjoy my surroundings when I travel by electric moped.

7. Habit
    7.1. I would need to make big changes to my travel pattern to start using shared (electric) bicycles or electric mopeds.
    7.2. I tend to use the same mode of transport when travelling.
    7.3. I tend to use the same route when travelling.
    7.4. I am open to trying new products and services.
    7.5. I am open to trying new digital applications.
8. Reliability
    8.1. I am confident that there will always be a shared vehicle available at the station.
    8.2. I am confident that there will always be a shared vehicle available for my return trip to the station.
    8.3. I am willing to pay more to have the certainty of having the shared vehicle for the entire round trip (leaving the station and coming back after the activity).
9. Perceived risk
    9.1. I feel safe when riding an electric moped.
    9.2. I feel safe when travelling by public transport in the Netherlands.
    9.3. I feel safe when riding an (electric) bicycle.
10. Sustainability
    10.1. I am concerned about the effects of climate change.
    10.2. I am aware of the impact transport has on climate change.
    10.3. I have adjusted my travel behaviour due to the impact it has on the climate.
11. Health
    11.1. I believe walking is a healthy way of travelling
    11.2. I believe cycling is a healthy way of travelling
    11.3. I believe that using electric vehicles (electric bicycle or moped) is a healthy way of travelling.
    11.4. I believe that using bus/tram/metro is a healthy way of travelling.
    11.5. I believe that using the train is a healthy way of travelling.
    11.6. I take health benefits of different modes into account when making travel choices.

## B. Exploratory factor analysis

We perform an exploratory factor analysis (EFA) on 48 attitudinal statements related to the technology acceptance UTAUT2 model (Venkatesh et al., 2012), tailored to the topic of SMM as an egress mode. The full detailed analysis is reported by Geržinič et al. (2025). To extract the factors, we apply the maximum likelihood method and the oblimin factor rotation (Schreiber, 2021). We then iteratively remove items that have a loading below 0.3 (Field, 2013), have a communality that is below 0.2 (Child, 2006) or a cross-loading that is either above 0.4 (Taherdoost, 2016) or more than 75% of the main factor loading (Samuels, 2017). The final model retains 25 items, loading onto eight factors. The KMO-value is 0.84 (meritorious), Bartlett's test is significant and the matrix determinant is acceptable ($1.13 \cdot 10^{-5}$). The final model is depicted in Table 7.

*Table 7. Final EFA model, with 25 items loading onto eight factors*

| Items | Factors | | | | | | | |
|---|---|---|---|---|---|---|---|---|
| | 1 | 2 | 3 | 4 | 5 | 6 | 7 | 8 |
| intention_1 | 0.937 | | | | | | | |
| intention_2 | 0.800 | | | | | | | |
| intention_3 | 0.850 | | | | | | | |
| intention_4 | 0.479 | | | | | | | |
| reliability_1 | | -0.699 | | | | | | |
| reliability_2 | | -0.977 | | | | | | |
| sustainability_1 | | | -0.882 | | | | | |
| sustainability_2 | | | -0.773 | | | | | |
| sustainability_3 | | | -0.668 | | | | | |
| social_1 | 0.655 | | | | | | | |
| social_3 | | | | 0.924 | | | | |
| social_4 | | | | 0.851 | | | | |
| effort_1 | | | | | 0.619 | | | |
| effort_3 | | | | | 0.862 | | | |
| effort_4 | | | | | 0.664 | | | |
| health_4 | | | | | | -0.850 | | |
| health_5 | | | | | | -0.768 | | |
| hedonic_2 | | | | | | | -0.832 | |
| hedonic_4 | | | | | | | -0.857 | |
| risk_1 | | | | | | | -0.635 | |
| facility_1 | | | | | *0.246* | | | 0.550 |
| facility_2 | | | | | | | | 0.605 |
| facility_3 | | | | | | | | 0.655 |
| facility_5 | | | | | | | | 0.722 |
| habit_5 | | | | | | | | 0.589 |

Next, for easier interpretation and further use of the factors, we give each factor a name based on the items loading onto it. Additionally, we inverse four of the factors (2, 3, 6, 7) to avoid using negative terms in factor naming. The final names of the factors are:

1. Intend to use SMM
2. Confident about SMM vehicle availability
3. Climate aware
4. SMM has a good societal image
5. SMM is easy to use
6. Using PT is a healthy way of travel
7. Mopeds are a fun and safe way of travel
8. Confident with using (digital) technology